\documentclass[]{pasj02}
\usepackage{threeparttable}
\usepackage[switch,mathlines]{lineno}

\Received{2025/07/09}
\Accepted{2026/01/20}
 
\begin{document}

\title{The Supernova Remnant G284.3$-$1.8 and Its Relation to the Gamma-ray Binary 1FGL~J1018.6$-$5856}
\author{
Natsuki \textsc{Terano},\altaffilmark{1}\altemailmark\email{m2321004@a.konan-u.ac.jp} 
Takaaki \textsc{Tanaka},\altaffilmark{1}\altemailmark\email{ttanaka@konan-u.ac.jp} 
Hiromasa \textsc{Suzuki},\altaffilmark{2}
Rei \textsc{Enokiya},\altaffilmark{3,4}
Hiroyuki \textsc{Uchida},\altaffilmark{5}
Kai \textsc{Matsunaga},\altaffilmark{5}
Takuto \textsc{Narita},\altaffilmark{5}
Yasuo \textsc{Fukui},\altaffilmark{6}
and 
Toshiki \textsc{Sato}\altaffilmark{7}}

\altaffiltext{1}{Department of Physics, Konan University, 8-9-1 Okamoto, Higashinada, Kobe, Hyogo 658-8501, Japan}
\altaffiltext{2}{Faculty of Engineering, University of Miyazaki, 1-1 Gakuen Kibanadai Nishi, Miyazaki, Miyazaki 889-2192, Japan}
\altaffiltext{3}{National Astronomical Observatory of Japan, 2-21-1 Osawa, Mitaka, Tokyo 181-8588, Japan}
\altaffiltext{4}{Faculty of Science and Engineering, Kyushu Sangyo University, 2-3-1 Matsukadai, Fukuoka 813-8503, Japan}
\altaffiltext{5}{Department of Physics, Graduate School of Science, Kyoto University, Kitashirakawa Oiwake-cho, Sakyo-ku, Kyoto 606-8502, Japan}
\altaffiltext{6}{Department of Physics, Nagoya University, Furo-cho, Chikusa-ku, Nagoya 464-8601, Japan}
\altaffiltext{7}{Department of Physics, School of Science and Technology, Meiji University, 1-1-1 Higashi Mita, Tama-ku, Kawasaki, Kanagawa 214-8571, Japan}


\maketitle

\begin{abstract}
G284.3$-$1.8 is a supernova remnant with a radio shell and thermal X-ray emission. Located near its center is the gamma-ray binary 1FGL J1018.6$-$5856, although the physical association 
between the two systems is not clear yet. 
Our X-ray spectroscopy with Suzaku reveals that  G284.3$-$1.8 and 1FGL J1018.6$-$5856 have compatible absorption column densities of $N_\mathrm{H} = 6\textrm{--}7 \times 10^{21}~\mathrm{cm}^{-2}$, indicating that the two systems have similar distances. 
The actual distance is determined as $3~\mathrm{kpc}$ using $\mathrm{^{12}CO}$ ($J=1\textrm{--}0$) data obtained with NANTEN. 
The X-ray spectrum of G284.3$-$1.8 shows a strong K-shell emission line of Mg, confirming that the earlier claim that the SNR is one of the few Mg-rich SNRs. 
Comparing recent stellar models taking into account the ``shell merger'' processes, we find that the obtained Mg-to-Ne mass ratio of $M_\mathrm{Mg}/M_\mathrm{Ne} = 0.73^{+0.07}_{-0.03}$ and Si-to-Mg mass ratio of $M_\mathrm{Si}/M_\mathrm{Mg} = 0.44\pm0.03$ suggest a supernova explosion that would have left a neutron star. 
The characteristics of 1FGL J1018.6$-$5856, on the other hand, are better explained with a model in which its compact object is neutron star. 
The present results, therefore, would suggest a possible scenario where G284.3$-$1.8 and 1FGL J1018.6$-$5856 are both remnants of a common supernova explosion although further observational tests are necessary.  
\end{abstract}

\section{Introduction}
At the end of their lives, stars more than ten times more massive than the sun completes nuclear fusion reactions and undergoes gravitational core collapse, causing supernovae (SNe). 
The properties of the progenitors are key ingredients in understanding the explosion mechanisms and the nature of compact objects left after the SNe. 
X-ray ejecta emission can provide us with the information on abundances of elements synthesized in the progenitors and during the SNe, and thus help us put constraints, for example, on the progenitor masses (e.g., \cite{Katsuda2018}).  

According to recent X-ray studies of Mg-rich supernova remnants (SNRs) such as N49B \citep{Sato2024} and G359.0$-$0.9 \citep{Matsunaga2024}, 
X-ray spectroscopy can work as a unique tool also to probe the activity inside the progenitor systems during pre-SN phases. 
\citet{Sato2024} and  \citet{Matsunaga2024} found high Mg-to-Ne ratios in the SNRs, and claimed such peculiar elemental abundances can nicely 
be explained if the ``shell merger'' processes \citep{Yadav2020,Andrassy2020} occurred inside the progenitors. 
Shell mergers are a process, in which the Ne- or O-burning is merged with the outer shell before the core collapse. 
The process substantially alter the elemental abundances synthesized in the star as well as its density structure.

The SNR G284.3$-$1.8 is another candidate Mg-rich SNR as reported by \citet{Williams2015}, who analyzed limited regions of the SNR with XMM-Newton. 
G284.3$-$1.8 has been observed at a wide range of wavelengths, including optical (e.g., \cite{van1973}), radio (e.g., \cite{Milne1989}) and X-ray (e.g., \cite{Williams2015}). 
Its age is estimated to be $\sim 10^4~\mathrm{yr}$ \citep{Ruiz1986}. 
Interestingly, at the center of G284.3$-$1.8 lies the gamma-ray binary 1FGL J1018.6$-$5856 although the connection between the two systems is not clear yet. 
1FGL J1018.6$-$5856 was discovered in the Fermi LAT data as a GeV gamma-ray source with flux and spectral variation with a 16.6-day period, which 
is attributed to orbital modulation \citep{Fermi2012}. 
Variable radio, X-ray \citep{Fermi2012}, and TeV gamma-ray \citep{HESS2012, HESS2015} counterparts were later found by follow-up observations.

We report here X-ray and radio ${}^{12}\mathrm{CO}$ 
line observations of the SNR G284.3$-$1.8, aiming to clarify its physical connection with the gamma-ray binary 1FGL J1018.6$-$5856. 
We discuss the connection based on distances estimated for the two systems, and based also on the progenitor mass, hence the compact remnant, of the SN of G284.3$-$1.8. 
All the statistical errors are quoted at $1\sigma$ confidence intervals.

\section{Observations \& Data Reduction}
We analyzed X-ray data of G284.3$-$1.8 obtained with Suzaku. Table~\ref{tab:log} summarizes the Suzaku observation log. 
In what follows, we analyzed data from the X-ray Imaging Spectrometer (XIS; \cite{Koyama2007}). 
The data reduction was performed using HEASoft version 6.32.1 with the calibration database released in October 2018. 
The XIS consists of four cameras with X-ray charge-coupled devices (CCDs) named XIS0, 1, 2, and 3. 
The CCDs of XIS0, 2 and 3 are front-illuminated (FI) devices whereas that of XIS1 is a back-illuminated (BI) CCD. 
We here analyze data from the FI CCDs.  
XIS2 and part of XIS0 suddenly showed an anomaly on November 2006 and June 2009, respectively, and thus we used (part of) XIS0 and XIS3 data. 
The non-X-ray background was constructed from the night Earth data using xisnxbgen \citep{Tawa2008}. 
Following the recipe provided by the XIS team, the cumulative flickering pixel maps were applied to the observation data. 
The redistribution matrix file (RMF) and ancillary response file (ARF) were generated with xisrmfgen and xissimarfgen, respectively \citep{Ishisaki2007}. 

We also analyzed $^{12}\mathrm{CO}$ ($J =1\textrm{--}0$) data from the NANTEN telescope.
The datasets were taken from the NANTEN Galactic Plane Survey (NGPS; \cite{Mizuno2004}).
The NGPS observations were carried out with $\sim1.1$ million data points in total, covering $l=200^\circ\textrm{--}60^\circ$ width in Galactic longitude and $|b|<20^\circ$ width in Galactic latitude.
The beam size, grid spacing, velocity resolution, and typical rms noise fluctuations are $2.6^\prime$, $4^\prime$, 0.65 $\mathrm{km~s^{-1}}$, and $\sim0.4$ K, respectively.

\begin{table}[h]
 \caption{Suzaku Observation Log}\label{tab:log}
   \begin{tabular}{cccc} \hline
   Obs. ID & Date & Start Time & Exposure [ks]  \\ \hline
   407069010 & 2012/06/20 & 22:19:00  & 73\\
407070010 & 2012/06/24 & 23:43:04  & 17\\
407071010 &2012/06/15 & 00:35:33  & 21\\
407096010 & 2012/06/27 & 17:05:53  & 60\\
504053010 & 2009/07/07 & 07:32:25  & 40\\
        \hline
   \end{tabular}
\end{table}

\section{Results}
\subsection{X-ray spectrum}
\begin{figure}[bt]
\begin{center}
\includegraphics[width=8cm]{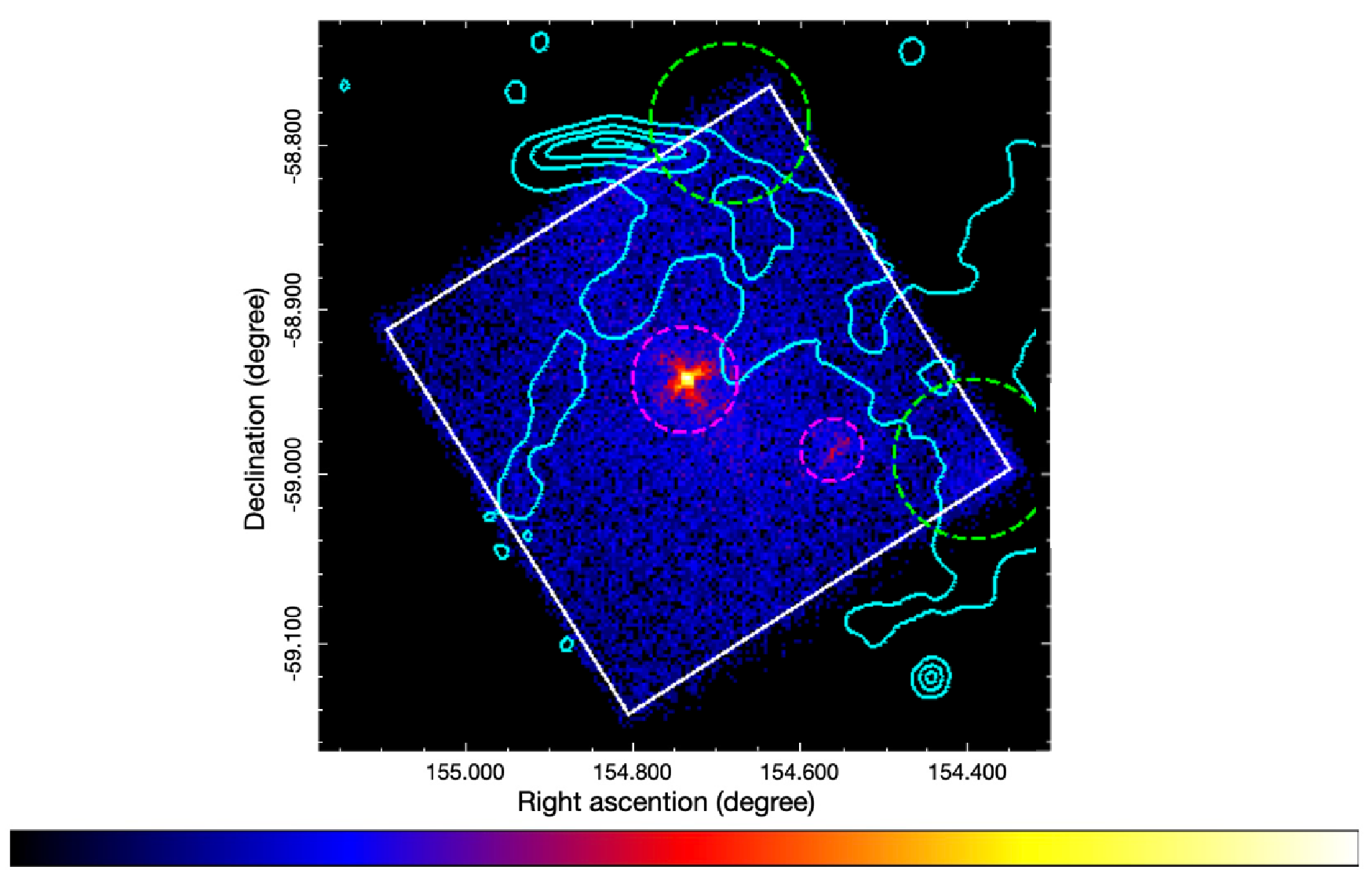} 
\end{center}
\caption{
Suzaku XIS3 image of G284.3$-$1.8 in the 0--12 keV energy band (color scale) from obsID 407069010,overlaid with the Molonglo Observatory Synthesis Telescope (MOST) radio contours at 843 MHz (cyan). The color bar shows X-ray surface brightness in the logarithmical scale. The source region is shown by a solid box. The $\atom{Fe}{}{55}$ calibration source (green dashed circles) and the emission region of the gamma-ray binary 1FGL J1018.6$-$5856 and another point-like source (magenta dashed circles) were excluded from the region. 
 {Alt text: X-ray image of the region toward the SNR G284.3$-$1.8 obtained with Suzaku XIS.}
}
\label{fig:xisimage}
\end{figure}

Figure~\ref{fig:xisimage} is the XIS image of the region. The point-like source located at the center of the field-of-view is the gamma-ray binary 1FGL~J1018.6$-$5856. 
Another point-like source is detected to the west of 1FGL~J1018.6$-$5856. 
The emission from G284.3$-$1.8 is almost uniform and seems to be spread over the field-of-view. 
We thus extracted spectra from the region excluding the two point-like sources as well as the $^{55}\mathrm{Fe}$ calibration sources. 
We co-added XIS0 and XIS3 spectra, and then combined all the spectra obtained from the observations listed in table~\ref{tab:log}. 
We use XSPEC (version 12.13.1) for spectral analysis described below.

Since the emission from G284.3$-$1.8 seems to occupy the whole field-of-view, we cannot extract background spectrum inside the field-of-view. 
We thus model the X-ray background at the same time as the emission from G284.3$-$1.8. 
Following \citet{Uchiyama2013}, we fitted the X-ray background with a model consisting of the Foreground Emission (FE), the Galactic Ridge X-ray Emission (GRXE), and the Cosmic X-ray Background (CXB). 
The FE is modeled as two-temperature plasma emission in the Collisional Ionization Equilibrium (CIE). 
The GRXE is composed of a two-temperature CIE component and the emission from the cold matter. 
The parameters for the FE and GRXE were fixed to those by \citet{Uchiyama2013} except for the absorption column density for the GRXE, and the normalizations of the FE and GRXE with fixing the 
ratios of low- and high-temperature components fixed to those by \citet{Uchiyama2013}. The CXB was modeled as a power law with a photon index of 1.4 and a 2--10~keV intensity of $6.38 \times 10^8~\mathrm{erg}~\mathrm{cm}^{-2}~\mathrm{s}^{-1}~\mathrm{sr}^{-1}$ by referring to \citet{Kushino2002}. 
The absorption column density for the CXB was set to twice that for the GRXE. 
We employed the NEI model for the emission from G284.3$-$1.8. 
We allowed the abundances of Ne, Mg, Si (= S), and Fe (= Ni) to vary. The other abundances cannot be constrained by the fit and thus are fixed to the solar values. 
We used Tuebingen--Boulder interstellar medium absorption model (TBabs; \cite{Wilms2000} for all the emission components described above. 
We referred to the abundance table by \citet{Wilms2000} for the solar abundances. 

\begin{figure}[bt]
\begin{center}
\includegraphics[width=8cm]{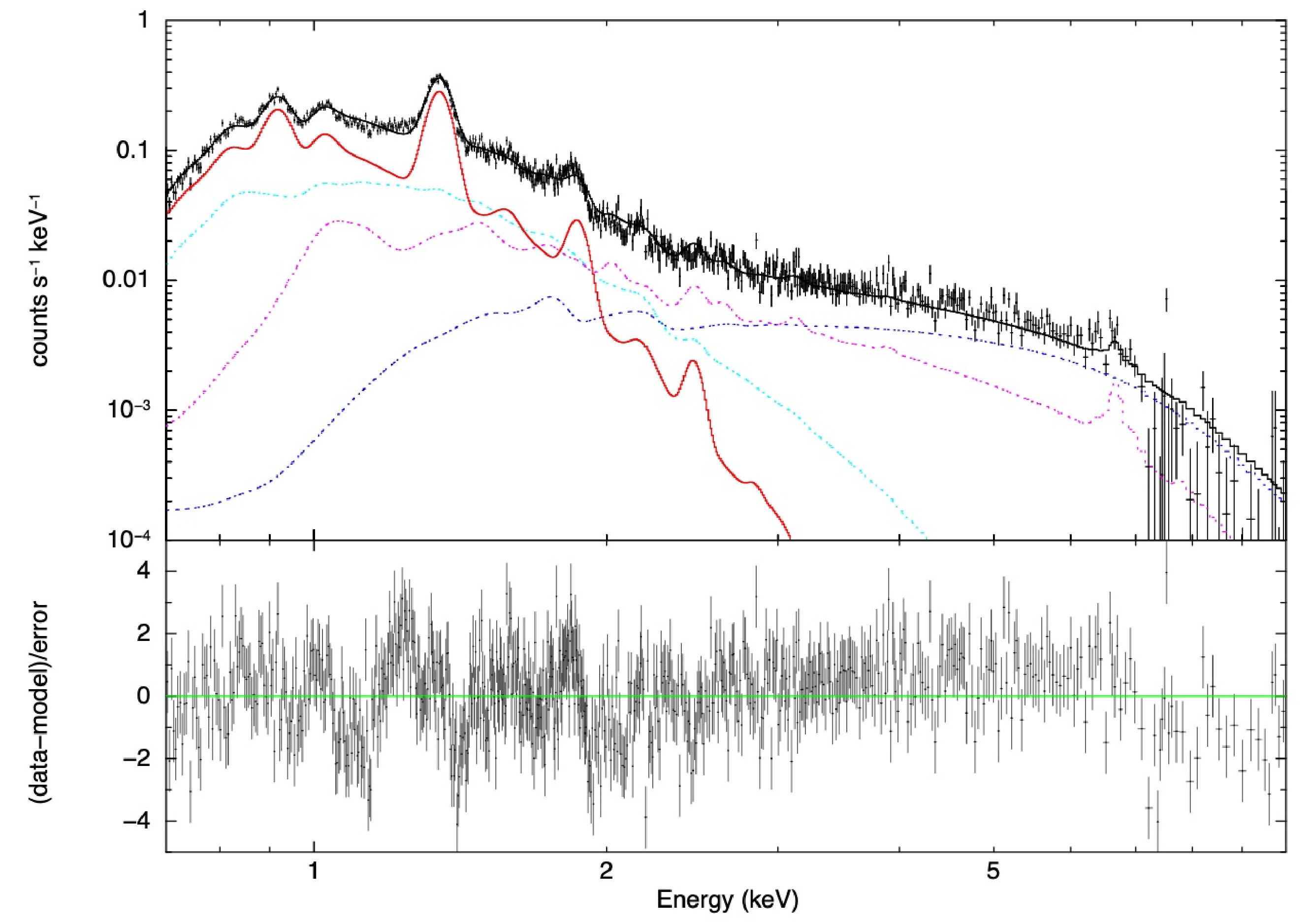} 
\end{center}
\caption{
NXB-subtracted Suzaku FI (XIS0+3) X-ray spectrum of G284.3$-$1.8 in the 0.7-10 keV band. The best-fit plasma model (red), and the background model (FE ; cyan, GRXE ; magenta, CXB ; blue).
{Alt text: X-ray spectrum of the SNR G284.3$-$1.8 in obtained with Suzaku XIS and the best-fit model.}
}
\label{fig:xisspec}
\end{figure}

The best-fit model is plotted with the spectrum in figure~\ref{fig:xisspec}, and the best-fit parameters are summarized in table~\ref{tab:para}. 
Only a lower limit of $\tau > 1.6 \times 10^{12}~\mathrm{{cm}^{-3}~s} $ was obtained for the ionization age, indicating that the SNR plasma is in CIE. 
The most characteristic feature of the present results is the significantly larger Mg abundance relative to Ne, confirming that G284.3$-$1.8 is a Mg-rich SNR as already pointed out by \citet{Williams2015}. Our best-fit parameters are somewhat different from what was obtained by \citet{Williams2015}, who analyzed spectra of small emission features in the SNR. 
To directly compare our analysis result to that of \citet{Williams2015}, we extracted spectra from a region similar to the ``west'' region used by \citet{Williams2015}. 
In fitting the spectra, we treated the X-ray background in the same manner as the above analysis, where we simultaneously fitted the GRXE component with the model by \citet{Uchiyama2013}. 
We found that the mass ratios $M_\mathrm{Mg}/M_\mathrm{Ne}$ and $M_\mathrm{Si}/M_\mathrm{Mg}$, which are mainly used in the below discussion, 
are consistent with each other. 
This demonstrates the robustness of the conclusion of the present work. 

\begin{table}[htp]
 \caption{Best-Fit Parameters}\label{tab:para}
\begin{threeparttable}
   \begin{tabular}{lll} \hline
   component & parameter & value  \\ \hline
    TBabs & $N_\mathrm{{H\_GRXE}}~ [10^{22}~\mathrm{cm}^{-2}]$ & $3.54^{+0.25}_{-0.19}$ \\
    vapec & $\mathrm{norm}$\tnote{1} ~$[10^{-2}]$ & $0.21\pm 0.01$ \\
    apec & $\mathrm{norm}$\tnote{1} ~$[10^{-2}]$ & $1.47^{+0.12}_{-0.05}$ \\ \hline
    TBabs &$N_\mathrm{{H\_G284}}~ [10^{22}~\mathrm{cm}^{-2}]$ & $0.67\pm0.03$  \\
    vnei & $kT$~$[\mathrm{keV}]$ & $0.317^{+0.006}_{-0.004}$  \\
     & $Z_\mathrm{Ne}$~$[\mathrm{solar}]$ & $0.93^{+0.26}_{-0.09}$  \\
     & $Z_\mathrm{Mg}$~$[\mathrm{solar}]$ & $2.82^{+0.93}_{-0.42}$  \\
     & $Z_\mathrm{Si}$~=~$Z_\mathrm{S}$~$[\mathrm{solar}]$ & $1.94^{+0.25}_{-0.18}$  \\
     & $Z_\mathrm{Fe}$~=~$Z_\mathrm{Ni}$~$[\mathrm{solar}]$ & $0.21^{+0.06}_{-0.02}$  \\
     & $Z_\mathrm{other}$~$[\mathrm{solar}]$ & $1~(\mathrm{fix})$  \\
      & $n_{e}t~[\mathrm{{cm}^{-3}s}]$ & $>1.6 \times 10^{12}$\\
      &    $\mathrm{norm}$\tnote{1}
      ~$[10^{-2}]$ & $1.53^{+0.28}_{-0.35}$ \\
    \hline
\end{tabular}\begin{tablenotes}\footnotesize
\item[1] Defined as $\frac {10^{-14}}{4\pi D^2} \int n_{\mathrm{e}}n_{\mathrm{H}} dV$, where $D$ is the distance to the source, $dV$ is the volume element, and $n_{\mathrm{e}}$ and $n_{\mathrm{H}}$ are the electron and hydrogen densities, respectively. The unit is cm$^{-5}$.
   \end{tablenotes}
\end{threeparttable}
\end{table}

\subsection{Molecular clouds toward G284.3$-$1.8}
\begin{figure}[bt]
\begin{center}
\includegraphics[width=8cm]{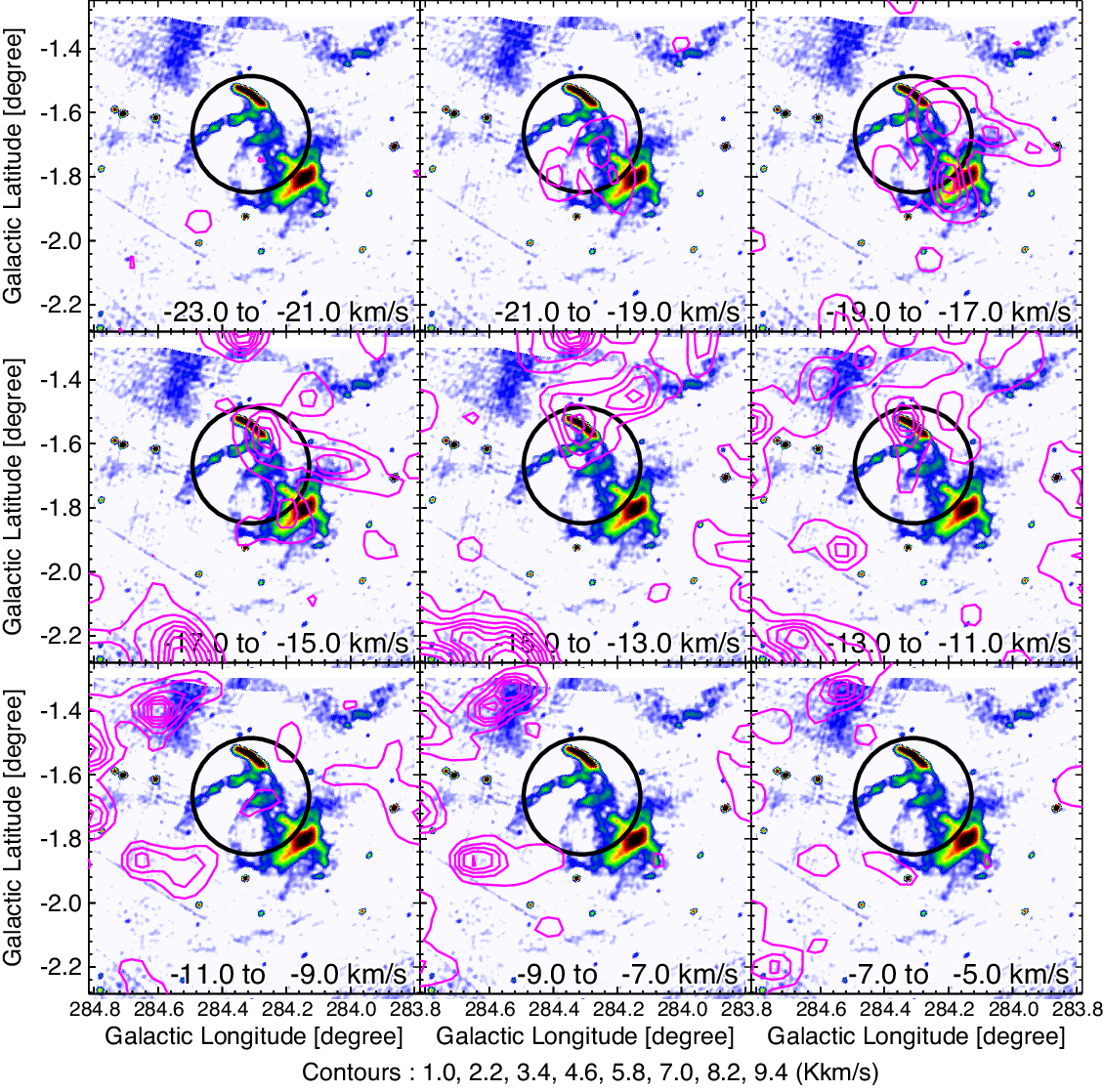}
\end{center}
\caption{
Velocity channel map of $\mathrm{^{12}CO}$ ($J$\textrm=1--0) emission obtained with NANTEN (magenta contours) overlaid on 843~MHz 
continuum emission obtained with MOST (color). The black circle indicates the rough location of the G284.3$-$1.8. 
{Alt text: Velocity channel map of the $\mathrm{^{12}CO}$ ($J$\textrm=1--0) line emission produced from the NANTEN data. The overlaid contours are 843~MHz 
continuum image with MOST.}
}
\label{fig:l_v}
\end{figure}

\begin{figure}[bt]
\begin{center}
\includegraphics[width=8cm]{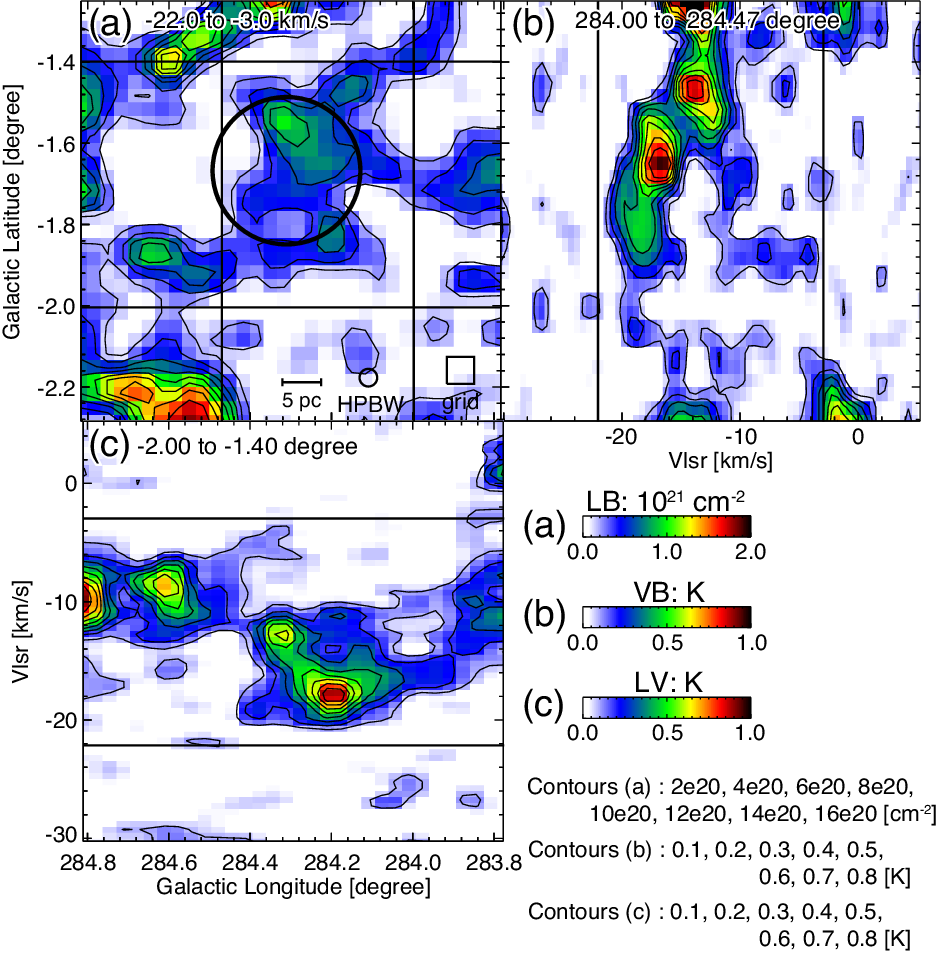}
\end{center}
\caption{
(a) Integrated intensity distribution of the associated cloud in $\mathrm{^{12}CO}$ ($J=1\textrm{--}0$) overlaid with black contours outlining $\mathrm{^{12}CO}$ ($J=1\textrm{--}0$) emission in the galactic coordinates. The black circle indicates the position of the G284.3$-$1.8. (b) Velocity-galactic latitude diagram of the associated cloud. (c) Galactic longitude-velocity diagram of the associated cloud. Only the color bar in (a) has a column density of molecular hydrogen. The unit of the color bar in (a) has been converted to the molecular hydrogen column density using the conversion factor derived by \citet{Okamoto2017}. 
{Alt text: Integrated intensity distribution of the $\mathrm{^{12}CO}$ ($J=1\textrm{--}0$) emission associated with the SNR G284.3$-$1.8 in the Galactic coordinate (a), velocity-Galactic latitude diagram (b), and Galactic longitude-velocity diagram (c). }
}
\label{fig:struct}
\end{figure}

We searched the NANTEN $\mathrm{^{12}CO}$ ($J=1\textrm{--}0$) data for signature of a molecular cloud associated with G284.3$-$1.8. 
Figure~\ref{fig:l_v} presents velocity channel map of $\mathrm{^{12}CO}$ ($J=1\textrm{--}0$) emission. 
We found a $\mathrm{^{12}CO}$ ($J=1\textrm{--}0$) emission that spatially coincides well with the shell of G284.3$-$1.8 in the velocity range from $-22~\mathrm{km~s}^{-1}$ to $-3~\mathrm{km~s}^{-1}$ (see figure~\ref{fig:struct}).

\section{Discussion}

\subsection{Estimation of distance}
\begin{figure}[bt]
    \begin{center} 
     \includegraphics[width=8cm]{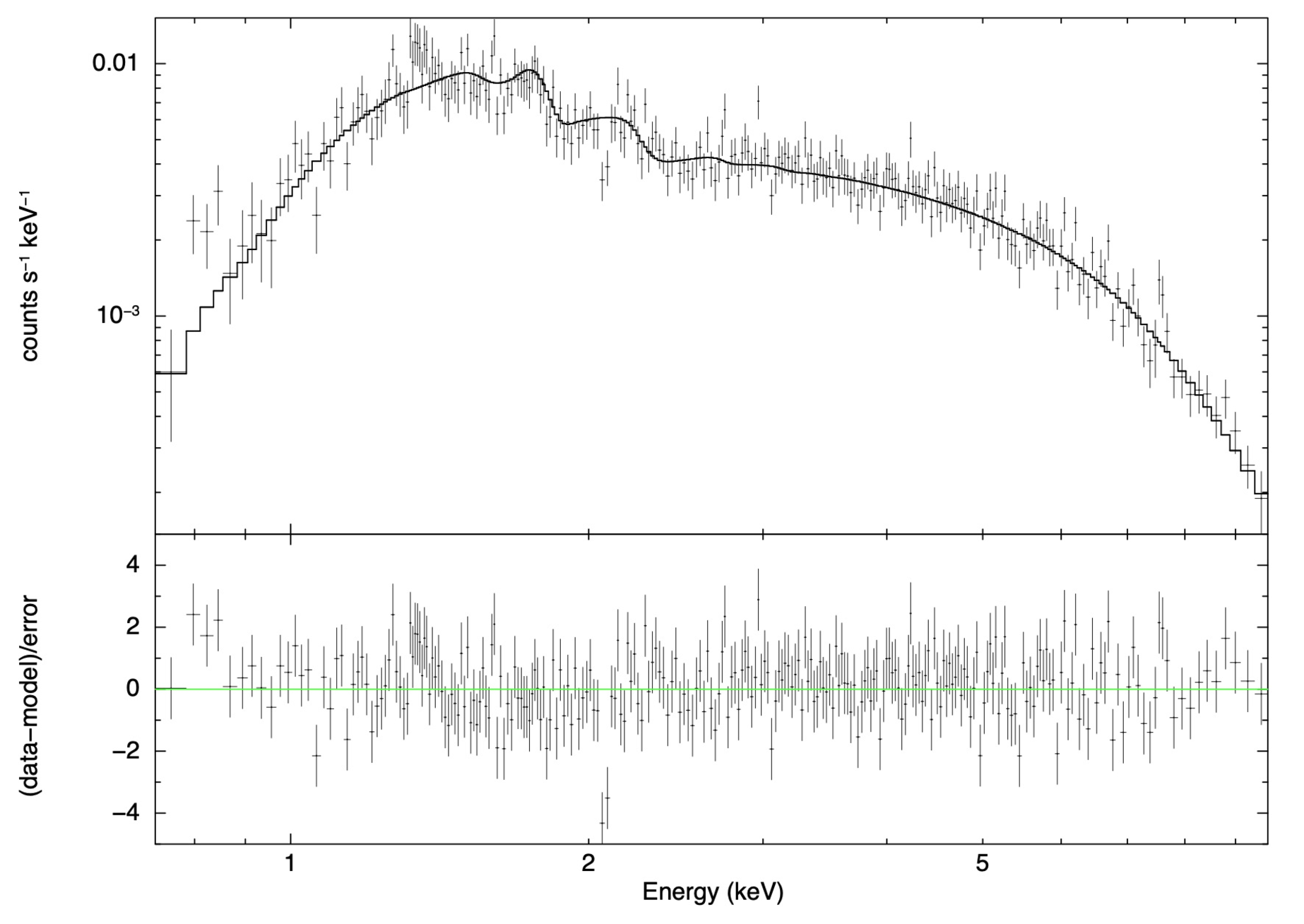}
    \end{center}
    \caption{NXB-subtracted Suzaku XIS (XIS0+3) X-ray spectrum of 1FGL~J1018.6$-$5856 in the 0.7-10 keV band. 
    {Alt text: X-ray spectrum of the gamma-ray binary 1FGL~J1018.6$-$5856  obtained with Suzaku XIS. }
    }    \label{fig:xisspec_2}
\end{figure}

The SNR G284.3$-$1.8 lies in the same line of sight as the gamma-ray binary 1FGL J1018.6$-$5856, suggesting a possibility that they are both remnants of a common supernova explosion. 
If the two systems indeed have the same origin, they should be located at the same distance and thus should have compatible X-ray absorption column densities ($N_\mathrm{H}$).  
In order to compare $N_\mathrm{H}$ of the two systems, we extracted the spectrum of 1FGL J1018.6$-$5856 from the region indicated in figure~\ref{fig:xisimage}. 
Figure~\ref{fig:xisspec_2} presents the spectrum plotted with the best-fit absorbed power law model. 
The absorption column density obtained is $N_\mathrm{H} = (6.4 \pm 0.2) \times 10^{21}~\mathrm{cm}^{-2}$, which agrees with that of G284.3$-$1.8 within the $1\sigma$ errors (see table~\ref{tab:para} for $N_\mathrm{H}$ of G284.3$-$1.8). 
We thus conclude that the two systems are located at comparable distances. 

Having determined that the distances are comparable, we estimate the actual distance to G284.3$-$1.8 based on the NANTEN $\mathrm{^{12}CO}$ ($J=1\textrm{--}0$) data. 
To constrain the distance to G284.3$-$1.8, we compared the radial velocity of the associated molecular cloud with those of the spiral arm structures of the Milky Way. 
By referring to the schematic diagram of the Galactic spiral arms by \citet{Reid2019}, we found that the line-of-sight velocity obtained in \S3.2 ($-22$ to $-3~\mathrm{km}~\mathrm{s}^{-1}$) matches 
the near-side of the Carina arm. 
The distance is then estimated to be $\sim 3$~kpc. 
Our estimate is compatible with that of \citet{Ruiz1986}, who analyzed $\mathrm{^{12}CO}$ ($J=1\textrm{--}0$) data obtained with the 1.2~m Columbia Millimeter-Wave Telescope, but closer than that of \citet{Shan2019} based on optical extinction measurements. 

Figure~\ref{fig:struct} illustrates the spatial and velocity structure of the molecular cloud associated with G284.3$-$1.8. 
A CO cavity structure (see figures~\ref{fig:struct}b and \ref{fig:struct}c) suggests that the gas has been evacuated by the supernova shock, indicating a strong interaction between the cloud and the SNR (e.g., \cite{Koo1991}).
The column density to the SNR is estimated to be $N_\mathrm{H} = 0.4\textrm{--}1.8 \times 10^{21}~\mathrm{cm}^{-2}$ (see figure~\ref{fig:struct}a). 

\begin{figure}[bt]
    \begin{center} 
     \includegraphics[width=7cm]{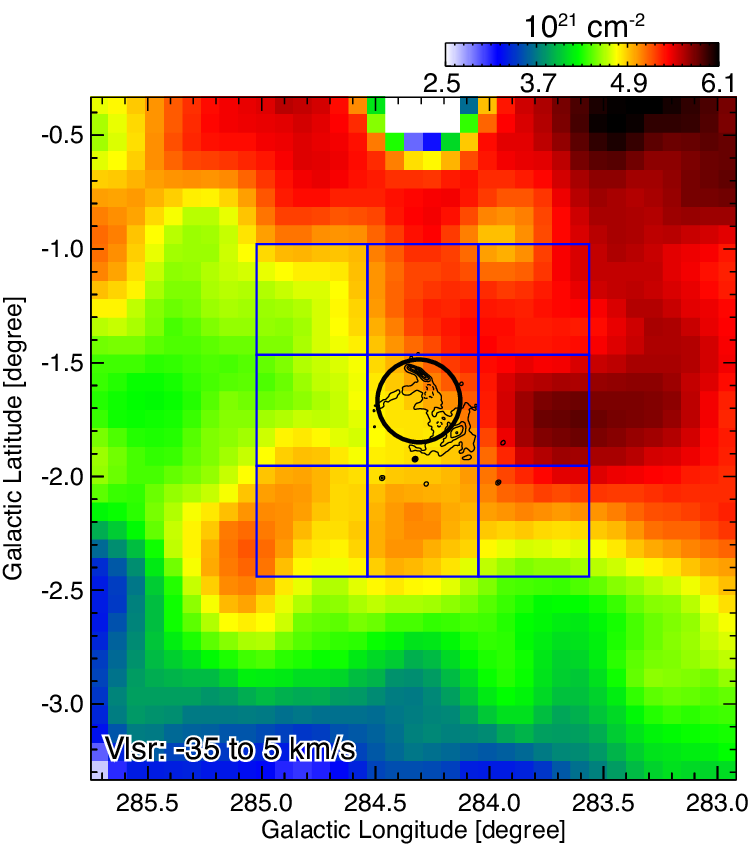}
    \end{center}
    \caption{Column density distribution of H\emissiontype{I} emission obtained with HI4PI in a velocity range from $V_\mathrm{lsr}=-35~\textrm{to}~5~\mathrm{km~s^{-1}}$. The black contours show the 843 MHz continuum map from MOST. The black circle indicates the rough location of G284.3$-$1.8. The regions used to derive the averaged H\emissiontype{I} spectra shown in figure~\ref{fig:abs_correlation} are indicated by the blue squares. 
    {Alt text: Spatial distribution of column density of  H\emissiontype{I} emission.}
    }    \label{fig:l_v2}
\end{figure}

To further investigate the interstellar medium associated with G284.3$-$1.8, we analyzed H\emissiontype{I} data from the HI4PI survey \citep{HI4PI2016}.
Figure~\ref{fig:l_v2} shows the H\emissiontype{I} distribution obtained by HI4PI. 
We found H\emissiontype{I} gas along the SNR shell at similar radial velocities to those of the CO clouds ($-35~\mathrm{km~s}^{-1}$ to $5~\mathrm{km~s}^{-1}$), 
indicating the presence of atomic gas associated with the SNR. 
The estimated H\emissiontype{I} column density in this velocity range is $N_\mathrm{H} = 5\textrm{--}6 \times 10^{21}~\mathrm{cm}^{-2}$.  
Then, the total (atomic and molecular) column density is $N_\mathrm{H} = 6\textrm{--}7 \times 10^{21}~\mathrm{cm}^{-2}$, which is consistent with the X-ray absorption column density.

\begin{figure}[bt]
\begin{center}
\includegraphics[width=8cm]{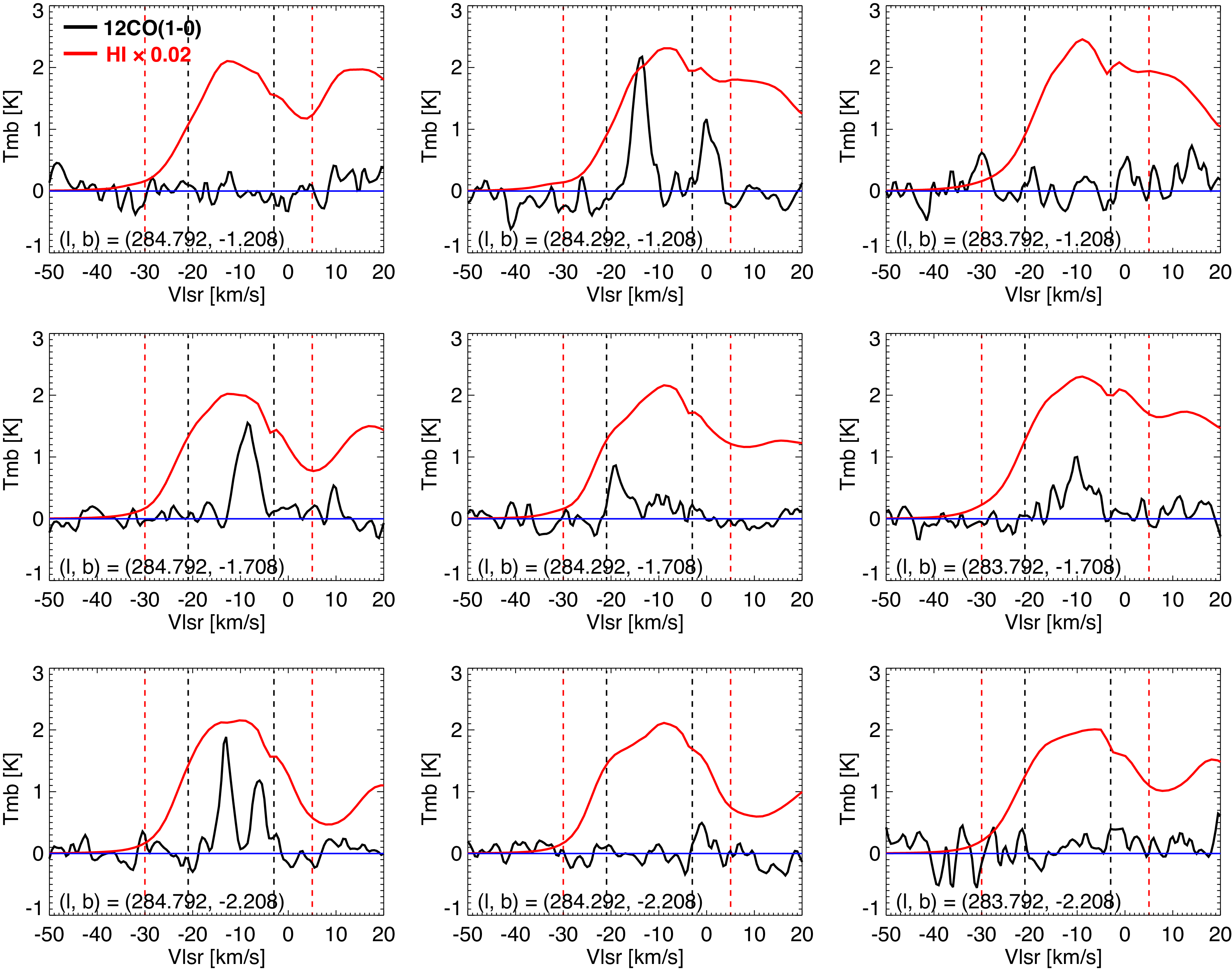}
\end{center}
\caption{$\mathrm{^{12}CO}$($J$\textrm=1--0) and H\emissiontype{I} spectral profiles within and surrounding eight regions of the remnant indicated in figure~\ref{fig:l_v2}. 
{Alt text: $\mathrm{^{12}CO}$($J$\textrm=1--0) and H\emissiontype{I} spectra extracted from the nine regions indicated in figure~\ref{fig:l_v2}. }
}
\label{fig:abs_correlation}
\end{figure}

In figure~\ref{fig:abs_correlation}, we show typical H\emissiontype{I} and CO profiles in G284.3$-$1.8 and its surrounding region.
We found clear H\emissiontype{I} absorption features within the remnant, while the surrounding regions show weaker absorption dips.
Higher angular resolution H\emissiontype{I} observations would be beneficial to confirm this connection in greater details.

In the study based on Gaia DR2 by \citet{Marcote2018}, 1FGL~J1018.6$-$5856 was estimated to be located at a distance of $6.4^{+1.7}_{-0.7}$ kpc, which does not agree with our estimate of  $\sim 3$~kpc. However, in Gaia DR3, the distance to the binary is determined to be $4.37 \pm 0.42~\mathrm{kpc}$ following the method described in \citet{Carretero2023} (Rib\'o, private communication). 
This would ease the tension between our conclusion and the Gaia distance to 1FGL~J1018.6$-$5856 although we need confirmation of the DR3 result in future works. 

\subsection{Progenitor of G284.3$-$1.8}
In \S3.1, we obtained elemental abundances of the ejecta of the SNR G284.3$-$1.8. Based on the results, we can now constrain the progenitor mass and thus the nature of the compact remnant left 
behind, paying particular attention to the high Mg abundance. 
The Mg-to-Ne mass ratio of G284.3$-$1.8 amounts to $M_\mathrm{Mg}/M_\mathrm{Ne} = 0.73^{+0.07}_{-0.03}$, which is significantly higher than the solar value, $M_\mathrm{Mg}/M_\mathrm{Ne} \sim 0.35$. 
As discussed on N49B by \citet{Sato2024} and on G359.0$-$0.9 by \citet{Matsunaga2024}, a plausible explanation of Mg-rich ejecta is the so-called shell merger process in the progenitor right before the core collapse, where Ne- or O-burning shells merge with outer layers (e.g., \cite{Yadav2020}).  Ne-burning shell intrusions promote processes such as ${}^{20}\mathrm{Ne}(\alpha,\ \gamma){}^{24}\mathrm{Mg}$, making $M_\mathrm{Mg}/M_\mathrm{Ne}$ higher. 
O-burning shell mergers, on the other hand, result also in higher $M_\mathrm{Si}/M_\mathrm{Ne}$ by enhancing processes such as ${}^{20}\mathrm{Ne}(\alpha,\ \gamma){}^{24}\mathrm{Mg}(\alpha,\ \gamma){}^{28}\mathrm{Si}$. 

\begin{figure}[bt]
\begin{center}
\includegraphics[width=9cm]{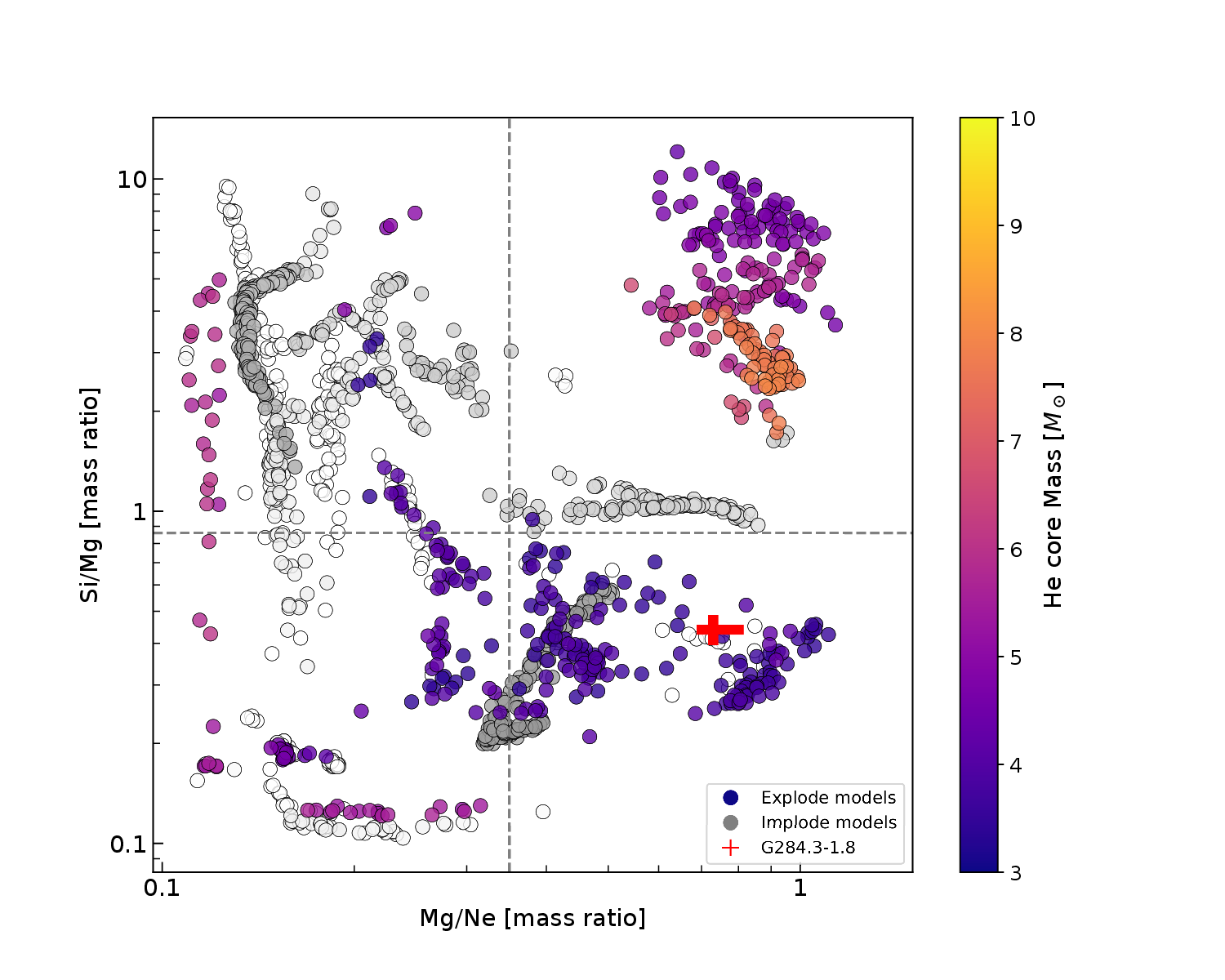}
\end{center}
\caption{Mass ratios $M_\mathrm{Mg}/M_\mathrm{Ne}$ and $M_\mathrm{Si}/M_\mathrm{Mg}$ in the O-rich layers of stellar models with $12\textrm{--}27~M_\odot$ taken from \citet{Sukhbold2018}, and distinction between exploding and non-exploding models based on \citet{Ertl2016} (See \cite{Sato2024} for details). The colored and gray circles represent explosion and implosion cases, respectively, with the gray dashed lines indicating the solar values. The color bar in correspond to He core mass. The red point indicate the ratios obtained from the present spectral analysis of G284.3$-$1.8. 
{Alt text: Relation between $M_\mathrm{Mg}/M_\mathrm{Ne}$ and  $M_\mathrm{Si}/M_\mathrm{Mg}$ of stellar models by \citet{Sukhbold2018} together with the ratios obtained for G284.3$-$1.8. }
}
\label{fig:distri}
\end{figure}

In figure~\ref{fig:distri}, we compare the observed mass ratios, $M_\mathrm{Mg}/M_\mathrm{Ne}$ and $M_\mathrm{Si}/M_\mathrm{Mg}$, 
with those of the theoretical models by \citet{Sukhbold2018}, which are also used by \citet{Matsunaga2024}. 
The explodability of each model is determined based on the criterion by \citet{Ertl2016}, following \citet{Sato2024}.
This figure clearly indicates that G284.3$-$1.8 belongs to the population with higher $M_\mathrm{Mg}/M_\mathrm{Ne}$ ratios. 
The population is divided into two groups depending on $M_\mathrm{Si}/M_\mathrm{Mg}$ ratios. 
Based on the relatively lower $M_\mathrm{Si}/M_\mathrm{Mg}$ of $0.44 \pm 0.03$, G284.3$-$1.8 is classified into the group with higher $M_\mathrm{Mg}/M_\mathrm{Ne}$ and lower $M_\mathrm{Si}/M_\mathrm{Mg}$, indicating that the progenitor experienced Ne-burning shell intrusion before the supernova explosion. 
There should be some uncertainties about the He core mass estimate indicated in figure~\ref{fig:distri} since differences in stellar evolution codes and stellar parameters 
(e.g., metallicity, overshoot, rotation, binary, nuclear reaction rates, etc.) could significantly affect the estimate. 
In any case, the Ne-burning shell intrusion likely to have occurred in the progenitor of G284.3$-$1.8 would have facilitate the explosion, according to the result by \citet{Sato2024}, who discussed the explodability of massive stars, including those that experienced the Ne-burning shell intrusion, based on the criterion by \citet{Ertl2016}. 
Such an explosion should have left a neutron star as the compact remnant \citep{Heger2003}.

The above discussion relies on the single-star models by \citet{Sukhbold2018}. 
The progenitor of G284.3$-$1.8, however, could have been a binary star as we explore such a possibility in what follows. 
According to the study by \citet{Laplace2021}, their binary-stripped star models can experience shell mergers when their He core mass is lower than $\sim 5M_\odot$. 
\citet{Laplace2021} also found that, compared to single stars with the same He-core mass, binary-stripped stars tend to have smaller compactness parameters \citep{Oconnor2011}, indicating that binary-stripped stars have higher explodability. 
Even in the binary-star case, the peculiar elemental abundances observed in G284.3$-$1.8 suggest that the progenitor had a relatively small He-core mass and 
left a neutron star after the supernova explosion. 

Another thing to note is the effect of explosive nucleosynthesis, which can, in principle, change the mass ratios discussed here in figure~\ref{fig:distri}. 
The small amount of Si in the observation would imply that the effects of explosive nucleosynthesis are small. 
 \citet{Sato2024} and \citet{Matsunaga2024} performed 1D supernova simulations to estimate the impact of the effect. 
 According to their results, explosive nucleosynthesis slightly (by $\sim 10$\%) decreases $M_\mathrm{Mg}/M_\mathrm{Ne}$ and increases $M_\mathrm{Si}/M_\mathrm{Mg}$ by a factor of $\sim 2\textrm{--}4$ in the case of stars with Ne-burning shell intrusion. Even if we take into account the effect, therefore, our conclusion should not be changed. 
 
We have found that the observed mass ratios, $M_\mathrm{Mg}/M_\mathrm{Ne}$ and  $M_\mathrm{Si}/M_\mathrm{Mg}$, imply a violent shell burning activity inside the progenitor, 
such as ``shell merger'', that would facilitate the explosion of a massive star. 
The characteristics of 1FGL J1018.6$-$5856, on the other hand, would better be explained if its compact object is a neutron star. 
Two major models have mainly been considered as acceleration/radiation mechanisms for gamma-ray binaries \citep{Mirabel2006}. 
One model, the so-called pulsar wind scenario, considers a young rotation-powered pulsar as the compact object in the binary system. 
In this model, particles (presumably electrons and positrons) are accelerated at the shock formed by collision of the relativistic wind from the pulsar and stellar wind from the companion star. 
The second model is the so-called micro-quasar scenario since it assumes a jet from the compact object, either a black hole or a neutron star, as the acceleration/radiation site.
As was discussed for a similar gamma-ray binary system, LS~5039, by \citet{Kishishita2009}, the relatively stable X-ray orbital modulation \citep{An2015} would better be reproduced with the first model, in which the compact object is a neutron star.  
Both G284.3$-$1.8 and 1FGL~J1018.6$-$5856, therefore, seem to contain neutron stars, suggesting a possibility of an interesting scenario where G284.3$-$1.8 and the compact object of 1FGL J1018.6$-$5856 are remnants of a common supernova explosion. 

The common-origin scenario for G284.3$-$1.8 and 1FGL~J1018.6$-$5856 still needs observational tests before one can confirm it. 
As discussed above, even with the DR3 result, we still have a tension between the Gaia distance to 1FGL~J1018.6$-$5856 and our estimated distance to G284.3$-$1.8 based on  the NANTEN $\mathrm{^{12}CO}$ ($J=1\textrm{--}0$) data. 
Another test is regarding the compact object in 1FGL~J1018.6$-$5856. 
Our X-ray spectroscopy of G284.3$-$1.8 implies that the compact remnant of the supernova explosion is a neutron star. 
On the other hand, the nature of the compact object in 1FGL~J1018.6$-$5856 is still under debate. 
The orbital parameters of the 1FGL~J1018.6$-$5856 system obtained through radial velocity measurements of the optical companion still allows 
the mass range of the compact object for a black hole (e.g., \cite{vanSoelen2022}). 
If detected, pulsations from 1FGL~J1018.6$-$5856 would give a decisive answer that the compact object is a neutron star.

\section{Conclusion}
We analyzed Suzaku XIS data of the SNR G284.3$-$1.8, at the center of which is the gamma-ray binary 1FGL J1018.6$-$5856. 
According to our spectroscopy, the two systems have compatible X-ray absorption column densities of $6\textrm{--}7 \times 10^{21}~\mathrm{cm}^{-2}$, indicating that they are at comparable distances. 
Analyzing $\mathrm{^{12}CO}$ ($J=1\textrm{--}0$) data obtained with the NANTEN telescope, we also found a molecular cloud associated with G284.3$-$1.8, whose distance was estimated to be 3~kpc. 
The Suzaku XIS spectrum of G284.3$-$1.8 indicates that the SNR is a Mg-rich supernova remnant, with a Mg/Ne mass ratio of $0.73^{+0.07}_{-0.03}$. 
We have shown that the supernova explosion would have left behind a neutron star by comparing the observed elemental abundances and stellar models.
On the other hand, the characteristics of 1FGL J1018.6$-$5856 are better explained if its compact object is a neutron star. 
Therefore, our result suggests a possible scenario where  G284.3$-$1.8 and 1FGL J1018.6$-$5856 are both remnants of a common supernova explosion 
although some observational tests are still necessary.


\end{document}